\documentstyle[12pt]{article}
\input psfig.tex
\begin{document}
\centerline{\Large\bf Angular Dependence of Neutrino Flux in KM$^3$ Detectors}
\centerline{\Large\bf
in Low Scale Gravity Models}

\bigskip
\centerline{\large\bf Pankaj Jain$^1$, Supriya Kar$^1$, Douglas W. McKay$^2$
}
\centerline{\large\bf
 Sukanta Panda$^1$ and John P. Ralston$^2$}

\bigskip
\begin{center}
$^1$Physics Department\\
I.I.T. Kanpur, India 208016\\
$^2$Department of Physics \& Astronomy\\
University of Kansas\\
Lawrence, KS 66045\\
\end{center}

{\bf Abstract:} Cubic kilometer neutrino telescopes are capable of
probing fundamental questions of ultra-high energy neutrino
interactions.  There is currently great interest in neutrino
interactions caused by low-scale, extra dimension models.  Above 1 PeV
the cross section in low scale gravity models rises well above the
total Standard Model cross section.  We assess the observability of
this effect in the 1 PeV - 1000 PeV energy range of kilometer-scale
detectors, emphasizing several new points that hinge on the enhancement 
of neutral current cross sections with respect to charged current cross
sections.  A major point is the importance of ``feed-down''
regeneration of upward neutrino flux, driven by new-physics neutral
current interactions in the flux evolution equations.  Feed-down is
far from negligible, and it is essential to include its effect.  We
then find that the angular distribution of events has high
discriminating value in separating models. In particular the
``up-to-down'' ratio between upward and downward-moving neutrino
fluxes is a practical diagnostic tool which can discriminate between
models in the near future.  The slope of the angular distribution, in
the region of maximum detected flux, is also substantially different
in low-scale gravity and the Standard Model.  These observables are
only weakly dependent on astrophysical flux uncertainties.  We
conclude that angular distributions can reveal a breakdown of the
Standard Model and probe the new physics beyond, as soon as data
become available.

\section{Introduction} The Ultra-high energy neutrino nucleon cross
section $\sigma_{\nu N}$ is a topic of fundamental physical
importance.  Low scale gravity models \cite{add,rs} predict
enhancement of the neutrino-nucleon neutral current type cross section
at center of mass energies above the fundamental gravity scale of
about 1 TeV \cite{NS99,DD99,JMPR}.\footnote{In the present context, by
neutral current type cross section we mean there is no leading,
charged lepton produced.  The shower is essentially hadronic, which
includes the case of black hole final states.}  Consequences for
cosmic ray physics have been studied in application to the highest
energy cosmic rays
\cite{NS99,DD99,JMPR,highest,cosmicBH,Emparan2,ring1,AFGS,ring2},
where there is great interest in possible violation of the Greisen,
Zatsepin, Kusmin (GZK) bound \cite{gzk}, at roughly 10 EeV ($10^{19}
eV$). There is also great interest in application to an intermediate
range \cite{JMPR,jaime,jaimefeng}, roughly 0.1 - 100 PeV ($10^{14} -
10^{17}$ eV).  The history and future of the highest energy
experiments \cite{highex} and intermediate energy experiments
\cite{intex} provide a tremendous impetus for these studies pointing
toward new physics.

Whatever the model, there are rich opportunities to study fundamental
high energy interactions by focusing on (1) the ratio of neutral
current type events to charged current events and (2) the {\it angular
distribution} of events in upcoming experiments.  The
neutral-to-charged ratio and the angular distribution shape do not
depend on the uncertainties of overall flux normalizations.  The
primary uncertainty in all cosmic ray comparisons with theory -- the
overall scale of the flux -- drops right out.  For a range of models
with standard and reasonable flux spectral indices, the angular
distributions are also remarkably {\it insensitive} to the details of
the model.  The strongest determining factor in the shape of the
angular distribution is the fundamental physics of the interaction
cross section itself.  We emphasize and explore this fact, showing
that arrays now planned or under construction could stringently test
the Standard Model and proposals for new physics simply on the basis
of the neutral-to-charge ratio and the {\it slope} of the angular
distribution of neutrino-nucleon events.  In the context of extra
space-time dimensions, data could determine or bound such details as
the scale and number of extra dimensions in the present models.

Here we continue earlier work \cite{JMPR} that examined possible
signatures of enhanced $\sigma_{\nu N}$ in kilometer scale detectors.
Extensions of the currently operating AMANDA \cite{amanda} and RICE
\cite{rice, rice2} experiments to ICECUBE \cite {icecube} dimensions would
certainly explore the 1 PeV - 100 PeV region.  In the case of RICE,
modest improvements even allow a reach above the EeV range.  In
\cite{JMPR} we used a linear extrapolation of the low scale gravity
mediated neutral current $\sigma_{\nu N}$ from the low energy, $\surd
s \ll 1$ TeV, region, to the $\surd s \geq 1$ TeV region and found a
very sharp suppression of the up-to-down ratio compared to the
standard model that set in at about 5 PeV for M = 1 TeV and at about
50 PeV for M = 2 TeV.

In that earlier study \cite{JMPR}, we did not apply other
extrapolations of the cross section to the up-to-down calculation, nor
did we include the ``feed down'' effect \cite{bgzr,Frichter}, which
results from neutral current interactions degrading higher energy
neutrinos as they travel through the earth and ``feeding'' the flux at
lower energies.  In the standard model, this effect is small above 1
PeV where the flux decrease steepens and the neutral current cross
section is too small to compensate.  In contrast, this effect turns
out to be extremely important after including the gravity induced
neutral current $\sigma_{\nu N}$, which rises rapidly with energy.
This point has not been explicitly recognized previously in
connection with gravity enhancement. 

A number of new gravity effects relevant to energies above the
fundamental scale, applicable to physics at the Large Hadron Collider
(LHC), kilometer cubed detectors ($KM^3$) and GZK energies, have been
proposed recently.  In both Arkani-Hamed, Dimopoulos and Dvali (ADD)
\cite{add} and Randall and Sundrum (RS) \cite{rs} models, eikonal
treatment of the effective low energy amplitude (used as the ``Born
term'' input) has been studied \cite{Emparan1} and applied to LHC
\cite{grw2} and GZK \cite{Emparan2} energies.  In string realizations
of the ADD framework \cite{AADD}, ``stringy'' cross sections, relevant
just below M, have been estimated \cite{cpp,stringy,d+e}, as has the
black hole formation cross section \cite{bh,cosmicBH}, which may be
relevant above M, including the GZK energy region.  We have
investigated all of these options, and find that the eikonalized ADD
low energy ``Born amplitude'' and ``geometrical'' black hole cross
sections lead to the largest and least model dependent effects in our
1 PeV to 1000 PeV $KM^3$ application.\footnote{We do not treat the
possibility of brane production and decay here \cite{branes}.}  Our
study goes beyond that reported recently in \cite{jaime} in two
respects: we emphasize higher energies and we include and analyze the
consequences of the new, eikonalized graviton exchange component of
the neutral current interaction.  This latter point also distinguishes
our work from a recent black hole detection study \cite{jaimefeng}.

We should note here that, though the neutral current does not produce
a prompt electron shower or leading muon characteristic of charged
current signatures, one expects that the hadronic shower, which is
completely electromagnetic after several radiation lengths, will be
an observable signature of neutral current interactions.
For this reason we regard all of the current detection mechanisms
to be relevant to our study, certainly including RICE, which can detect
a radio pulse from any kind of shower, AMANDA and ICECUBE.

\subsection{The Cross Section} At C.M. energies well above the Planck mass,
the classical gravity Schwarzschild radius $R_{s}(\surd s)$ is the
dominant physical scale.  The classical impact parameter,b, may make
sense in this domain, and the eikonal approximation be valid, for
values of b larger than $R_{s}$.  We will sketch the eikonal set-up
shortly. At smaller impact parameters, the parton-level geometrical cross
section
\begin{equation}
\hat\sigma_{\rm BH} \approx \pi r_S^2
\label{sigma_BH}
\end{equation}
provides a classical, static estimate of the cross section to form
black holes.  In Eq. \ref{sigma_BH} $r_S$ is the Schwarzchild radius of a $4+n$
dimensional black hole of mass $M_{\rm BH} = \sqrt{\hat s}$,
\begin{equation}
r_S = {1\over M}\left[M_{\rm BH}\over M\right]^{1\over 1+n}
\left[{2^n\pi^{(n-3)/2}\Gamma\left({3+n\over 2}\right)\over 2+n}\right]^{1\over 1+n} 
\end{equation}
where $\sqrt{\hat s}$ is the parton-parton or, in our case,
neutrino- parton C.M. energy, and $M$ is the 4+n-dimensional scale
of quantum gravity.\footnote{We use the mass scale convention
discussed in \cite{AFGS}, referred to as $M_{D}$ there.}  The black
hole production process is expected to give a dominant contribution when
$\sqrt{\hat s}>> M$.  Black holes will form only if the impact
parameter $b < r_S$.  To convert Eq.1 into an estimate for the neutrino-
nucleon cross-section, we fold it with with the sum over parton 
distribution functions and integrate over x-values, where $\hat s = xs$,
at a momentum transfer typical of the black hole production process:
\begin{equation} 
\sigma_{\nu N\rightarrow BH}(s) = \Sigma_{i}\int_{x_{min}}^{1}dx
\hat\sigma_{BH}(xs) f_{i}(x,q).
\end{equation}
Black hole formation requires $x>M^{2}/s$, so we take
$x_{min}=M^{2}/s$.  In addition, $q^{-1}=b<r_{S}$ is required.  We
adopt $q = \surd \hat s$ up to $\surd \hat s = 10 TeV$, the maximum
range in q of the CTEQ parton distribution functions \cite{CTEQ4}, the
set we use, and $q = 10 TeV$ when $\surd \hat s$ is above this value.
As remarked in \cite{AFGS}, the dependence of $\sigma_{\nu
N\rightarrow BH}(s)$ on the choice of $x_{min}$ and the treatment of q
is rather mild.

In the case of ADD model the black hole
production cross sections can be large for $n>2$, in which case the
fundamental scale can be of the order of 1 TeV.  A number of authors
have adopted this estimate and applied it to ultra relativistic,
parton level scattering. The approximation has been challenged on the
basis that quantum corrections should lead to exponential suppression
of individual channels, such as the black hole formation final state
\cite{voloshin,ralstonUNPUB}, with several, independent arguments
advanced in each case.  In defense of the ``black disk''
approximation, several authors also point to success of internal
consistency checks of the classical picture \cite{d+e,giddings,grw2}.  
Recent phenomenological studies seem to
be agnostic on this issue \cite{ring1,AFGS,ring2,jaimefeng,rizzo},
treating the phenomenological consequences of both versions.

In a string picture with scale $M_{s} < M$, there is a range of energy
$M_{s} \simeq \surd s$ where string resonances dominate
\cite{cpp}, and a range $M_{s} < \surd s \leq M$, where
stringball formation \cite{d+e} could dominate.\footnote{In this discussion
we suppress the $\hat s$ notation for convenience, though parton level
processes are intended.} The cross section can be roughly expressed as
 \cite{cpp}
\begin{equation}
\sigma_{SR}(\surd s) \sim g_{s}^2 \delta(s-M_{SR}^2), \surd s\simeq M_{SR},
\end{equation}
for the string resonance case.
Here $g_{s}$ is the (weak) string coupling constant and $M_{SR}$ is the
mass of a string resonance state. Similarly, for the stringball case,
estimates of the cross section give \cite{d+e}
\begin{equation}
\sigma_{SB}(\surd s) \sim 1/M_{s}^2, M_{s}/g_{s} <  \surd s <  M_{s}/g_{s}^2,
\end{equation}
where $M$ is a few times less than $M_{s}/g_{s}^2$ for weak coupling.
The impact of these various processes on the physics to be expected at the
LHC, at a next linear collider (NLC) and very large hadron collider (VLHC)
has been surveyed in a number of papers, summarized and referenced in
the Snowmass 2001 report of the extra dimensions subgroup \cite{bounds}.

In our application to $KM^3$ physics in this paper, we mentioned above
that the ``classical'' eikonal cross section \cite{thooft,Emparan1}
and the geometric black hole formation cross section are the only
cases where we find potentially observable effects.  We outline our treatments of the eikonal model in the
ADD \cite{add} and RS1 \cite{rs} pictures next.  The black hole cross
section needs no further elaboration.  The fundamental mass scale $M$
in the case of the ADD model for $n > 2$, can be of the order of 1
TeV, though new astrophysics analyses may constrain n = 3 more severely,
as we comment below \cite{hewett+spiropulu}. 
Similarly, in the RS picture the effective scale of gravity on
the physical brane, the lowest K-K mode mass, can be arranged to be of
the order of 1 TeV.  In all of our quantitative work, we set the scale
$M$ the same for every value of n we use in our comparisons.  As noted
earlier, the scale $M$ is the same as $M_{D}$ defined in \cite{grw2}
and discussed in \cite{AFGS}.

In the RS1 model with one extra dimension or in the ADD
model with several, a possible choice for the input amplitude to the eikonal
approximation, referred to as the Born amplitude, is given by,
\begin{equation}
i{\cal M}_{\rm Born} = \sum_i{ics^2\over M^2}{1\over q^2 + m_i^2}  
\end{equation}
where $c$ is the gravitational coupling strength, which is 
effectively Newtonian for ADD and electroweak for RS. 
Here $q=\sqrt{-t}$ is the momentum transfer. In the
Randall-Sundrum case, the sum runs over the massive K-K modes,
constrained to start at or above the TeV scale when c is of order
electroweak strength.  Their spacing is then also of TeV order.
In the ADD case, the index {\it i} must include the mass degeneracy
for the {\it i}th K-K mode mass value.  The spectrum is so nearly
continuous that an integral evaluation of the sum is valid, but must
be cut off at a scale generally taken to be of the order of M.
Taking the transverse Fourier transform of the Born amplitude, we get the
eikonal phase as a function of impact parameter, b,
\begin{equation}
\chi(s,b) = {i\over 2s}\int {d^2q\over 4\pi^2} \exp(i{\bf q}\cdot {\bf b})
i{\cal M}_{\rm Born}
\end{equation} 
For the ADD model, where $c = (M/\overline M_{P})^{2}$ and $\overline
M_{P} = 2.4\times 10^{18}$ GeV is the reduced, four dimensional Planck mass,

\begin{eqnarray}
\chi(s,b) & = & -{s (2^{2n-3}\pi^{{3n\over 2} -1})\over M^{n+2}\Gamma(n/2)} 
2\int_{0}^{\infty}dmm^{n-1}K_{0}(mb)\nonumber\\
          & = & \left({b_{c}\over b}\right)^n,
\end{eqnarray}
where
\begin{equation}
b_c^n = {1\over 2}(4\pi)^{{n\over 2}-1}\Gamma\left[{n\over 2}\right]
{s\over M^{2+n}}\ . 
\end{equation}
Because of the exponential decrease of $K_{0}$, the phase integral is
actually ultraviolet finite.
For the RS model, the corresponding expression is
\begin{equation}
\chi(s,b) =\sum_i{cs\over 2M^2}K_{0}(m_ib).
\end{equation}
For widely spaced KK modes in the Randall-Sundrum case, the lowest
few modes dominate and contribute insignificantly in the $1\ {\rm PeV}
< E_{\nu} < 100$ PeV region.  We do not discuss RS further.

The eikonal amplitude is then given in terms of the eikonal phase by
\begin{eqnarray}
{\cal M} &=& -2is \int d^{2}b\exp(i{\bf q}\cdot {\bf b})\left[\exp(i\chi)
 -1\right]\nonumber\\
          &=& -i4\pi s \int db bJ_0(qb)\left[\exp(i\chi)-1\right],
\end{eqnarray}
The eikonal amplitude for the case of ADD model can be obtained
analytically \cite{Emparan1,Emparan2,grw2} 
in the strong coupling $qb_c >>1$ and weak coupling
$qb_c<<1$ regimes.

In the strong coupling regime the eikonal amplitude can be computed using
the stationary phase approximation and is given by,
\begin{equation}
{\cal M} = A_n e^{i\phi_n}\left[{s\over qM}\right]^{n+2\over n+1}\ ,
\end{equation}
where 
\begin{equation}
A_n = {(4\pi)^{3n\over 2(n+1)}\over \sqrt{n+1}}\left[\Gamma\left({n\over 2
}+1\right)\right]^{1\over 1+n}\ ,
\end{equation}
\begin{equation}
\phi_n = {\pi\over 2} + (n+1)\left[{b_c\over b_s}\right]^n
\end{equation} 
and
$b_s = b_c(qb_c/n)^{-1/(n+1)}$.
In the weak coupling regime, $q\rightarrow 0$ the amplitude is
given by
\begin{equation}
{\cal M}(q=0) = 2\pi isb_c^2\Gamma\left(1-{2\over n}\right) e^{-i\pi/n}\ .
\end{equation}
As it turns out, the small q region contributes little to the cross section,
and we use the simple rule that the amplitude is set to its value at
$q=1/b_{c}$ for values $q\leq1/b_{c}$.

\begin{figure}[t,b]
\bigskip
\hbox{\hspace{0em}
\hbox{\psfig{figure=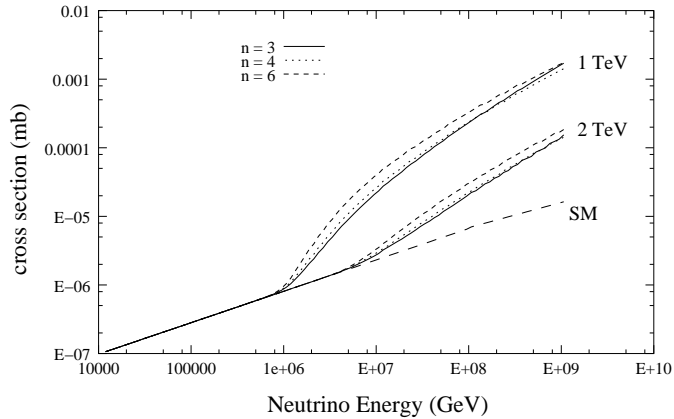,height=6cm}}}
\caption{The
neutrino-proton cross section $\sigma_{\nu p}$ in the
ADD model using the number of extra dimensions $n=3,4,6$
as a function of the neutrino energy $E_\nu$ for $E_\nu<10^8$ GeV.
The geometric black hole production cross section is included.
The solid, dotted and short
dashed curves correspond to $n=3,4$ and 6 respectively.
The long dashed curve represents the Standard
Model prediction.
}
\label{cross2}
\end{figure}

The parton-level cross section is calculated by assuming that it is
given by the Born term as long as $\hat s<M^2$. For $\hat s>M^2$ the
cross section is estimated by the eikonal amplitude.\footnote{A
discussion of the reliability of the eikonal amplitude in the $ 5\
{\rm TeV} \leq \surd s \leq 15$ TeV range is given in \cite{grw2}.}
For M = 1 TeV, for example, the actual matching between the Born and
eikonal amplitudes occurs in the range $\surd \hat s$ = 1 - 3 TeV,
depending on n and the value of $y = q^2/\hat s$.  In any case the
region $\surd\hat s\sim M$ contributes negligibly to the
cross-section, so the precise matching choice makes no difference in
the final result.  The eikonal calculation is not expected to be
reliable if the momentum transfer $q>M$. At large momentum transfer we
assume that the black hole production dominates the cross section. The
eikonal cross section is, therefore, cut off once the momentum
transfer $q > 1/r_S$.  The neutrino-parton differential cross-section
is folded with the CTEQ parton distributions and integrated over x and
y variables, consistent with our momentum transfer restriction on the
eikonal amplitude.  The CTEQ limit at $x =10^{-5}$ is exceeded only at
the high end of the energy range we study.  A standard power law
extrapolation is used when x does range below this value, though the
$\hat s$ values are so low that the contribution of this range to the
cross section is negligible.  In Fig. \ref{cross2} we plot the total
neutrino-nucleon cross sections for several different values of the
fundamental scale $M$ and the number of extra dimensions $n$,
including both eikonal and black hole production contributions.  The
cross section is clearly more sensitive to the choice of the scale
parameter, M, than to the number of dimensions, n.  In fact the
sensitivity to choice of n comes primarily through the dependence of
the differences in bounds on M corresponding to different choices of
n.  The strongest bounds on M come from astrophysical and cosmological
considerations, which unavoidably require some degree of modeling.

A recent review of experimental and observational constraints is given
in \cite{hewett+spiropulu}, where lower bounds on M are quoted for n =
3 from various analyses such as supernova cooling, regarded as the
least model dependent bound ($M \geq 2.5 TeV$), post-inflation
re-heating ($M \geq 20 TeV$), and neutron star heat excess ($M \geq 60
TeV$).  For n = 4, the most severe constraint is $M \geq 5 TeV$ in the
case of the post-inflation re-heating limit.  There are essentially no
constraints on the cases n = 5 and 6.  Laboratory lower bounds are
typically of order 1 TeV or less for all $n\geq2$, with LEP II
providing the strongest bound at 1.45 TeV for n=2.
\section{Event Rates of Downward Neutrinos} 
\begin{figure}[t,b]
\bigskip
\hbox{\hspace{0em}
\hbox{\psfig{figure=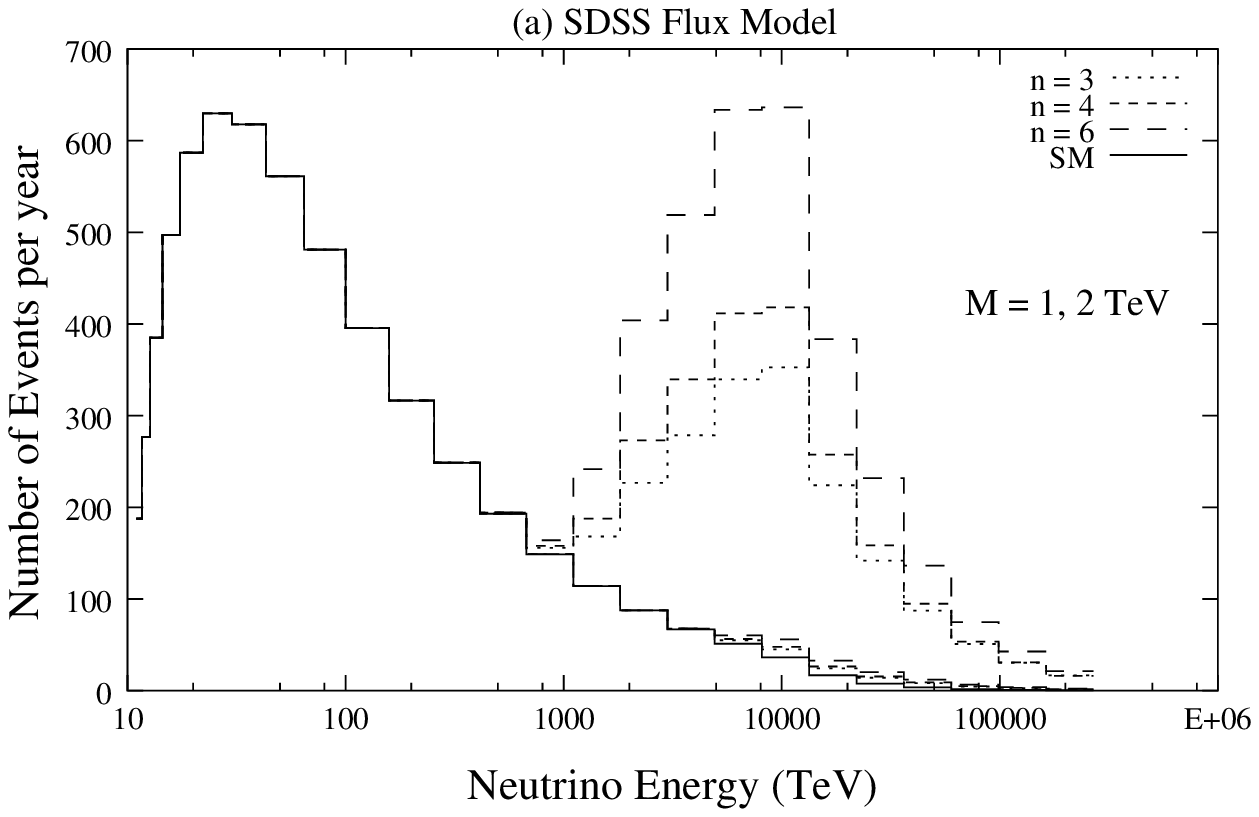,height=6cm}}}
\vskip 0.1in
\hbox{\hspace{0em}
\hbox{\psfig{figure=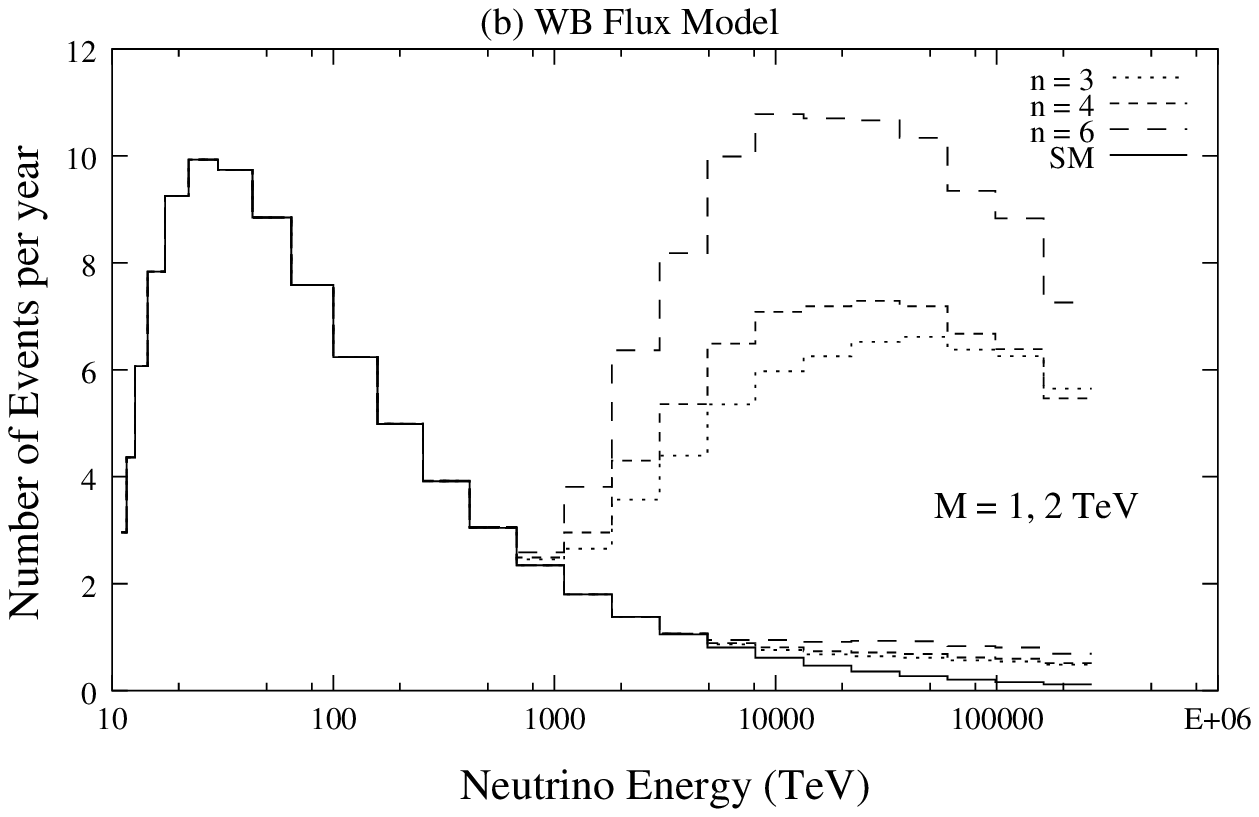,height=6cm}}}
\caption{The downwards event rate per cubic kilometer per year within
the Standard Model (SM) and in the ADD model for the fundamental 
scale $M=1, 2$ TeV and for the number of extra dimensions, $n=3,4,6$
using (a) the SDSS flux model and (b) the WB flux model.
The solid line shows the Standard Model prediction alone. The long dashed,
dotted and short dashed curves show the predictions of the neutral
current graviton exchange within the ADD model plus geometric, black
hole cross sections, and including the SM contribution. 
The upper and lower histograms correspond to the $M=1,\ 2$ TeV choices 
respectively. The new physics contributions are insignificant below 1 PeV,
where their contribution above the SM is not visible on this scale.
}
\label{event_rate}
\end{figure}

In Fig. \ref{event_rate}, we
show the downward event rate, defined as the number of interactions
from down- coming neutrinos, within a one kilometer cubed volume.  We
use two, quite different input flux assumptions to show the dependence
of rate on the flux.  A simple parameterization of an optically thick
source model of flux above 1 PeV given in Ref. \cite{sdss} (SDSS) is
used, along with the flux bound for optically thin source environments
of Ref. \cite{wb} (WB).  The flux in \cite{sdss} is about two orders
of magnitude larger than the bound in \cite{wb}, and it has roughly an
$E^{-2}$ power law behavior from 0.1 PeV to 10 PeV, and then it steepens
to approximately $E^{-3}$. We parametrized the SDSS flux such that it
falls as $E^{-2}$ for 10 TeV $<E<$ 10 PeV and as $E^{-3}$
for $E>$ 10 PeV. The WB bound, on the other hand, falls as
$E^{-2}$ over the whole energy range. The two flux curves cross at
about $10^3$ PeV.

Our estimate is made by taking the vertical flux result, multiplying
it by $2 \pi$ steradians and by the probability that a neutrino would
interact within 1 km in ice, with density 0.93 gm/cc, given the cross
section model in question.  To get an actual event rate for a given
detector, we would have to multiply by the acceptance of the detector.
In the SDSS flux model, Fig. 2a, the number of interactions peaks at
around 10 PeV for the $M=1$ TeV case, with encouragingly large
numbers of interactions, in the 350-650 range, induced by low-scale
gravity.  This is 10-20 times the SM interaction rate at the same
energy.  The number and the location of the peak rate depend upon the
flux model of course \cite{Frichter}.  This dependence is illustrated
in Fig. 2b, where the event rate for the WB bound is shown.  The event
rates are lower by a factor of about 50 and the gravity induced events
have a much broader peak, centered at about 15 PeV, compared to the
SDSS case. The peak is broader for the WB flux than for SDSS because
the WB limit falls as $E^{-2}$ throughout this energy region, while
the SDSS flux falls as $E^{-3}$ above 10 PeV, cutting off the higher
energy events more rapidly.  The shape of the SM event rate is the
same in both cases, since the flux shapes are the same below 10 PeV.

\begin{figure}[t,b]
\bigskip
\hbox{\hspace{0em}
\hbox{\psfig{figure=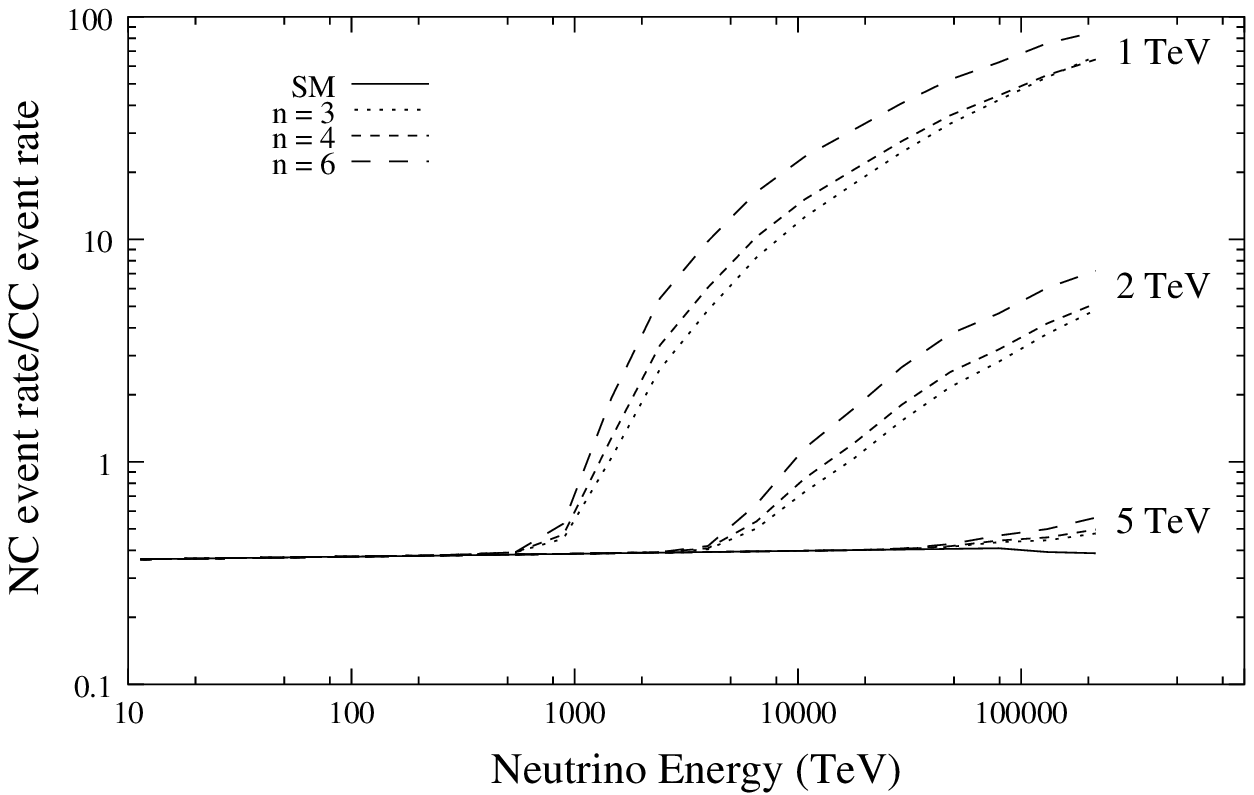,height=6cm}}}
\caption{The ratio of neutral current type events to SM charged current events
as a function of neutrino energy for n = 3, 4 and 6 and for M = 1,2, and 5 TeV.
Only the downward events are included in this plot. The neutral current type
interactions, in the sense used here, 
 are dominated by the black hole and eikonal components of the low scale
gravity amplitude above the scale of gravity.
}
\label{nc/cc}
\end{figure}

The excess above the SM is roughly half neutral current type events
arising from eikonalized graviton exchange and half black hole events,
which decay predominantly into hadrons.  Therefore an optimal
detection scheme requires sensitivity to the hadronic shower from the
deposited energy in the ice.  The ICECUBE and RICE detectors, for
example would respond readily to hadron showers at these energies.
The special role of enhanced neutral current type events, those with
hadronic signatures like the graviton exchange and black hole events,
prompts us to propose that the ratio of neutral current to charged
current events provides a powerful tool to uncover new interactions.
In particular, if the neutral current type component, distinguished by
hadronic dominated showers and no leading charged lepton, {\it
dominates}, as it does in the black hole and the eikonal regimes of
the low scale gravity models, there would be {\it no standard model
explanation}.  The ratio of neutral current type events to charged
current events, distinguished by the presence of a leading charged
lepton, is shown in Fig.\ref{nc/cc}.  The rapid rise that sets in
above the threshold for new physics is remarkable, and would be so
even without the black hole contribution.  Even scales $M > 2 TeV$ are
discernible with this observable.  Putting together the information
from several techniques, RICE and ICECUBE for example, one might well
separate the ``neutral'' versus ``charged'' character of events and
find a clear window on new physics.

\section{The Regeneration and Angular Dependence of Neutrino Flux} We
next calculate the up over down ratio of neutrino flux.  The downward
$\phi_0(E_\nu)$ is the flux of neutrinos incident on the surface of
the earth from the sky. We will consider only the diffuse neutrino
flux and assume that it is isotropic. Our calculations are easily
generalized if the flux is found to be non-isotropic or if we are
interested in individual sources.  The upward flux $\phi_{\rm
up}(E_\nu)$ is defined as the flux of neutrinos coming upwards from
the surface of the earth. This is the angular average
\begin{equation}
\phi_{\rm up}(E_\nu) = {1\over 2\pi}\int_{0}^{2\pi}d\Phi\int_{0}^{\pi/2}
\sin(\theta)d\theta \phi(E_\nu,\theta)
\end{equation}
of the flux $\phi(E_\nu,\theta)$ emerging from the earth, where
$\theta$ is the polar angle with respect to the nadir and $\Phi$
is the azimuthal angle. The ratio
$R$ is essentially the ratio of up-to-down event rates.  The event
rates are given by the product of flux, cross section, number density,
volume and acceptance, and to the extent that the latter four factors cancel
in the ratio, only the up-to-down ratio of fluxes survives.  This ratio
is affected by the energy dependence of the flux, but not to its
overall normalization.  The upward flux depends on the cross sections
of course, as we elaborate next.

In order to determine $\phi(E_\nu,\theta)$ we first need to solve the
{\it evolution equation} for the neutrino propagating through the
interior of the earth.  In the case of the standard model, the neutrino
cross section is dominated by the charged current, the neutral current
does little to evolve, or ``feed down'', the rapidly falling flux
above 1 PeV, and neutrinos basically get lost after collision inside
earth.  In the present case, on the other hand, the eikonalized
graviton exchange gives a large contribution to the feed down above 1
PeV, and we have to include this important enhancement of regeneration
of lower energy neutrinos from higher energy.\footnote{The black hole
production component, in contrast, essentially leads to loss of
neutrinos upon collision.  In the present context, the black hole
interaction acts like the charged current in the standard model.  The
eikonal component has the character of the usual neutral current,
transferring a small fraction of the neutrino energy to the hadron and
leaving a leading neutrino in the final state.}
\footnote{We do not treat the regeneration of $\tau$ neutrinos
through $\tau$ production by $\nu_{\tau}$ and subsequent $\tau$ decay
back to $\nu_{\tau}$\cite{beacom}.}
The large cross sections mean that neutrinos experience 
many interactions as they proceed through the earth, even
at angles near the horizon. For instance, just 5 degrees below the horizon
a 100 PeV neutrino traverses more than 20 interaction lengths of earth
before reaching the detector region.  This behavior supports our use of the
continuous evolution model, summarized below in Eq.17, since fluctuations
are small for UHE application, where the cross section is large.
The neutrino loses little energy during any particular collision and hence
it is reasonable to assume that it practically moves in a straight line path.
The evolution equation for the neutrino is given by \cite{bgzr,Frichter}
\begin{eqnarray}
{d\ln \phi\over dt}(E_\nu,\theta) &=& -\sigma_{W+BH}^{\nu N}(E_\nu)
-\sigma_{Z+E}^{\nu N}(E_\nu) + \int_{E_\nu}^\infty dE_\nu^\prime
{\phi(E_\nu^\prime,\theta)\over \phi(E_\nu,\theta)}{d\sigma_{Z+E}^{\nu N}\over
dE_\nu^\prime}(E_\nu^\prime,E_\nu)\nonumber\\
&\equiv & -\sigma^{\nu N}_{\rm eff}(E_\nu,\theta),
\label{evol_eqn}
\end{eqnarray}
where $\sigma_{W+BH}^{\nu N}$ is the ``neutrino absorbing'' W-exchange +
black hole cross section and $\sigma_{Z+E}^{\nu N}$ is the ``neutrino
regenerating'' Z-exchange + eikonalized graviton exchange cross section. 
In Eq.17, $dt=n(r)dz$ and $n(r)$ is the number density of nucleons at
any distance $r$ from the center of earth, radius $R_{e}$. 
Expressing the flux $\phi(E_\nu,\theta)$ as
\begin{equation}
\phi(E_\nu,\theta) = \phi_0(E_\nu)\exp[-\sigma_{\rm eff}(E_\nu,\theta)
t(\theta)],
\label{evol_soln}
\end{equation}
where the column density at upcoming angle of entry $\theta$, chord length
$2R_{e}cos\theta$,  is given by
\begin{equation}
t(\theta) = \int_{0}^{2R_{e}\cos\theta}n(z,\theta)dz, 
\end{equation}
we solve equations \ref{evol_eqn} and \ref{evol_soln} numerically by 
iteratively improving the flux $\phi(E_\nu,\theta)$.
Using this solution we can determine the ratio $R$ of up to down flux, 
\begin{equation}
R = {\phi_{\rm up}(E_\nu) \over \phi_0(E_\nu)} 
\label{ratio_defn}
\end{equation}
By using $R$, one sharply reduces the effects of experimental
systematics and flux normalization and isolates the dependence on
cross sections and flux shape.  Refinements of angular binning can add
information according to the size of the data sample, as we discuss below.

\begin{figure}[t,b]
\bigskip
\hbox{\hspace{0em}
\hbox{\psfig{figure=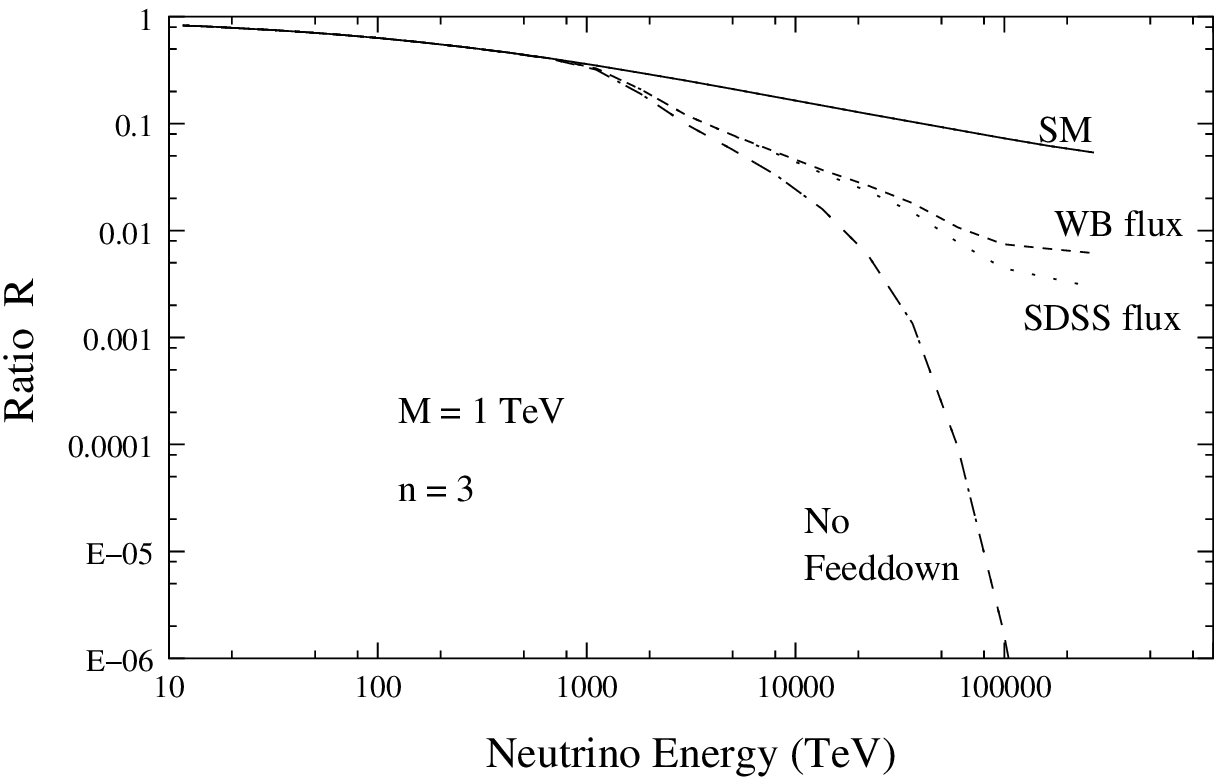,height=6cm}}}
\caption{The contribution of the regeneration term to the ratio $R$ of
the upwards and downwards flux within the ADD plus geometric black
hole production model with the fundamental scale $M=1$ TeV and for
the number of extra dimensions equal to 3. The long dashed line (flux
from \cite{sdss}) and short dashed line (flux from \cite{wb}) predict
nearly the same R as a function of energy for the full calculation,
including the regeneration term.  Ignoring the regeneration of flux,
one gets the lowest curve, identical for both flux models. The regeneration
term has a profound effect on
the up-to-down flux ratio R due to the large, gravity driven 
neutral current cross section.  Standard model prediction, the solid
line, is also shown and it is the same for both flux assumptions.  }
\label{evolution}
\end{figure}

In Fig. \ref{evolution} we plot the ratio $R$ as a function of the
neutrino energy.  This figure illustrates two points: the ratio $R$ is
insensitive to the normalization of the flux assumed and, given a
flux, the feed-down from higher to lower energies as the neutrinos
pass through the earth is a powerful effect.  In this application, our
two, quite different input flux assumptions, show the {\it
insensitivity} of R to the flux used.  The long dashed curve gives the
ratio for the larger flux with a ``knee'' from \cite{sdss}, while the
shorter dashed line gives the result of the smaller flux bound with
uniform $E^{-2}$ fall-off of \cite{wb}.  The solid curve refers to the
SM cross section input, the two flux models give indistinguishable
results in this case and are shown as one line. The lowest curve shows
the ratio as a function of energy with pure absorption, in the sense
that only the first two terms in $\sigma_{eff}$, Eq.17., are included.
The two flux models give the same result, as they should. The sum of
the eikonal and black hole cross section is used, the eikonal
providing the neutral-current cross section and the black hole the
purely absorptive cross section, as described in footnote 6.  The
cross section are computed in the ADD model, and the value $n=3$ and
the scale of quantum gravity $M=1$ TeV.  We see that the regeneration
term gives a big effect for $M = 1$ TeV, producing factors of more
than 10 above 20 PeV, and it is essential to include it in assessing
the consequences of the low scale gravity models. Though radically
different from each other, the two flux assumptions lead to nearly
identical values of R out to 50 PeV and then differ only weakly above
that, where the effect of the steeper decrease of SDSS flux shows up.

\begin{figure}[t,b]
\bigskip
\hbox{\hspace{0em}
\hbox{\psfig{figure=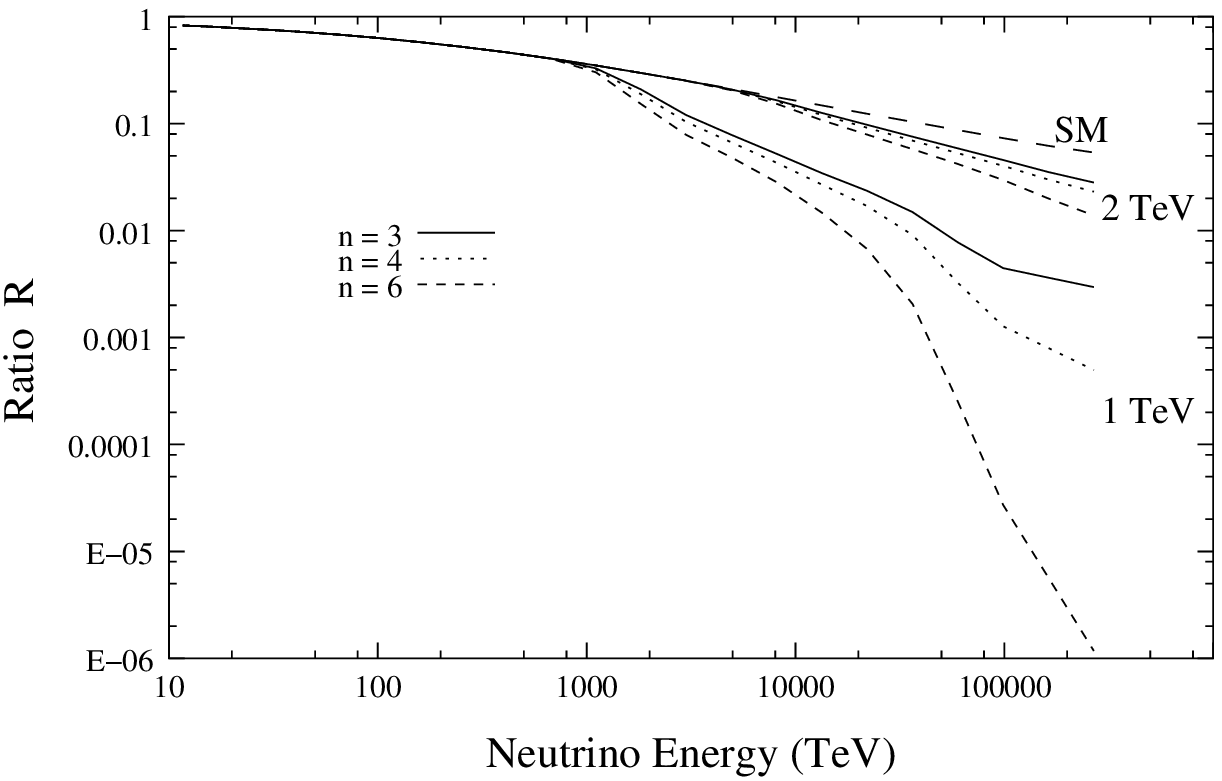,height=6cm}}}
\caption{The ratio $R$ as a function of neutrino energy for the 
fundamental scale $M=1, 2$ TeV and for the number of extra dimensions
3 (solid curves), 4 (dotted curves) and 6 (dashed curves). The 
Standard model (SM) prediction is also shown.
}
\label{ratio}
\end{figure}

In Fig. \ref{ratio} we plot the results for the ratio $R$ in the ADD
model as a function of the neutrino energy for the choice of the
fundamental scale $M=1,2$ TeV and the number of extra dimensions equal
to 3,4 and 6. The result depends on both the number of extra
dimensions and, especially strongly, on the fundamental scale $M$. If
the fundamental scale is larger than about 2 TeV, the KM$^3$ neutrino
detectors will not be able to distinguish the Standard model result
from the predictions of quantum gravity by using $R$ as a diagnostic.
However for $M\approx 1$ TeV, we find that the effect is very large
and, given enough flux, could be seen in these detectors.  For
energies above 1 PeV, a noticeable difference in $R$ between the two
cases is seen.  With sufficient data, a distinction between no
deviation from the standard model and a deviation corresponding to
$M=1$ with n = 3, 4 or 6 may be drawn.

\begin{figure}[t,b]
\bigskip
\hbox{\hspace{0em}
\hbox{\psfig{figure=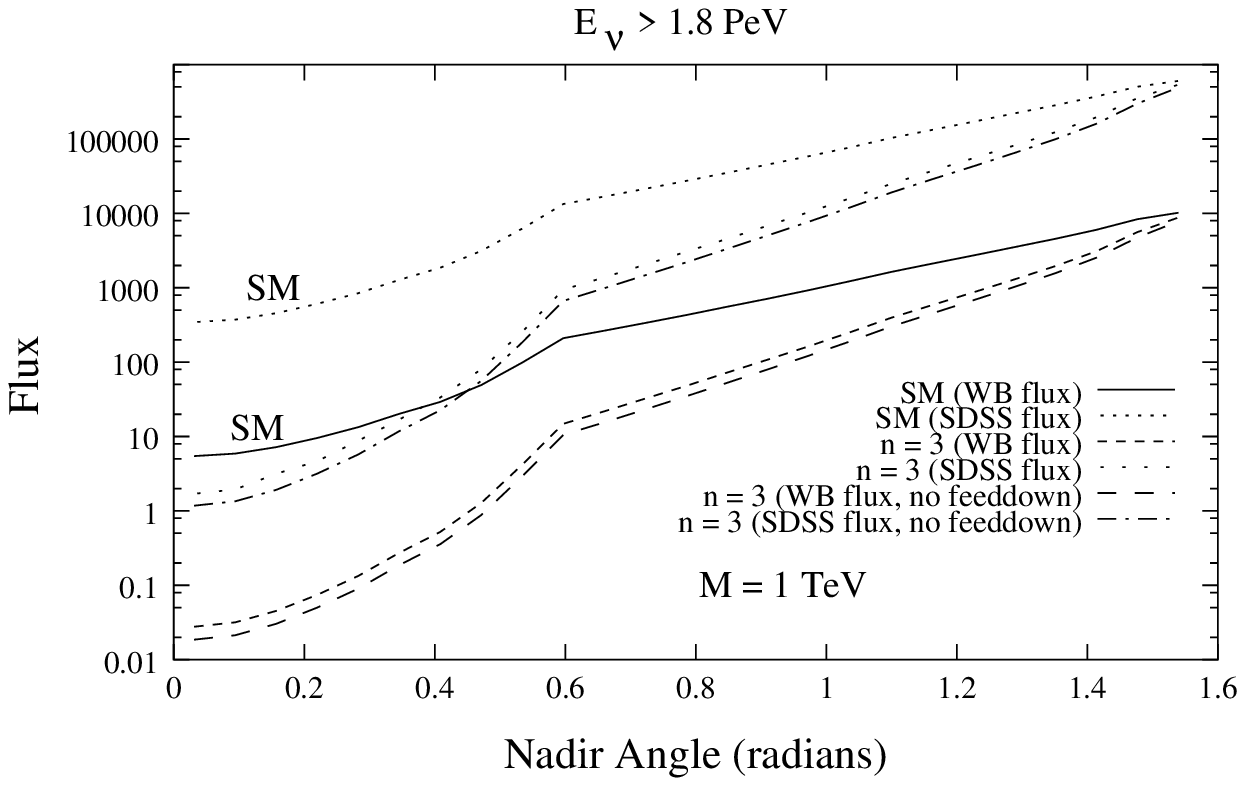,height=6cm}}}
\caption{The angular dependence of the upwards flux for the fundamental 
scale $M=1, 2$ TeV and for the number of extra dimensions, $n=3,4,6$.
The Standard model (SM) prediction is also shown.
Only neutrinos with energy $E_\nu>1.8$ PeV are considered. The $M=2$
TeV results are indistinguishable from the Standard Model, but the R value 
is different because downward event rate is larger in the M=2 low scale
gravity model (see Fig.\ref{event_rate}).
}
\label{flux_ang}
\end{figure}

The angular distribution of upward flux is an important diagnostic
tool, as we stressed in the Introduction.  In Fig. \ref{flux_ang}, we
show the integrated flux per square kilometer per year above $1.8$ PeV
as a function of nadir angle for $M=1,2$ and n=3,4,6. The (higher) 2
TeV curve is essentially the same as for the SM, while the (lower)
curves for $M=1$ show clear suppression at all nadir angles up to $\pi
/2$.  It is somewhat disguised by the log scale, but most of the
contribution to the ratio $R$ comes from the highest event rates near
the horizontal.  With several bins of good statistics data above 1
radian in nadir angle, one can distinguish the slopes of the flux
vs. nadir angle near the horizontal.  The difference between these
slopes for the SM and low scale gravity models when $M \simeq 1 TeV$
provides another new physics discriminator.  As in the case of the R
observable, the slope is insensitive to the flux {\it value},
enhancing its power at identifying the nature of the neutrino {\it
interactions}.

Realistically, one needs enough upcoming events at a given energy to
calculate a meaningful ratio.  The situation for the two flux models we
are discussing is shown in the table.

\begin{table}[h]
\begin{tabular}{|l|l|l|l|l|l|r|}                                         
\hline
\multicolumn{2}{|c|}{} &
\multicolumn{4}{c|}{Number of upward events per year} \\
\hline
n &$M$ & 1-10 PeV & 5 - 10 PeV & $>$ 10 PeV
 & $>$ 15 PeV \\ \hline
  3 & 1 & 83 & 7.2 & 1.3 & 0.22 \\
  4 & 1 & 79 & 5.4 & 0.62 & 0.07 \\ 
  6 & 1 & 74  & 2.6 & 0.09 & 0.004 \\  
  3 & 2 & 80 & 8.9  & 4.3  & 1.8  \\
  4 & 2 & 80 & 8.9  & 4.2  & 1.7  \\
  6 & 2 & 80 & 9  & 4.1  & 1.6  \\ \hline
  \multicolumn{2}{|c|}{SM}   & 80 & 8.8  & 4.2  & 1.7  \\
\hline
\end{tabular}
\caption{The numbers of upward events per year expected over the energy
ranges shown, for models with different n and $M$ choices and for the standard
model. The flux used is our rough approximation to SDSS above 1 PeV }
\end{table} 

For our two power law approximation to the flux of \cite{sdss} (SDSS),
we see from the table that there are enough events with $E_{\nu}> 10$ PeV in
a 10 year run to determine the ratio R.  Can one discriminate among
the different physics models for the cross section with the events?
Fig. \ref{ratio} shows that if there are enough upcoming events to
calculate a meaningful value for R, there are clear distinctions among
the models.  For example, taking a naive, purely statistics estimate,
the $M = 1$ TeV, $E_{\nu}>10$ PeV case would produce $13\pm3$, $6\pm2$
and $1\pm1$ events for n = 3, 4 and 6.  Combining these values with the
downward event rates (see Fig. 2), we see that the distinction between
the n = 3 case and the n= 6 cases is significant.  When $M = 2$ TeV,
the numbers of upcoming events is essentially the same for the low
scale gravity models and the SM. In 5 - 10 years of data, there would
be enough events to determine an up-to-down ratio reasonably well.
Figure \ref{event_rate} shows that the number of downward events is
larger in the low scale gravity models than in the SM, because of
their larger cross sections, so R values are different for the
different cases.  Thus, even for $M = 2$, values of R above 15 PeV are
distinctly different in the two classes of models, as one sees in
Fig. 5.  In fact, within the low scale gravity models themselves there
is more than a factor two difference between n = 3 and n = 6.
Moreover, the low scale gravity models can all be easily distinguished
from the SM with 5 years of data, with the number of events per year
shown in the table.  Even the cut $>15$ PeV allows the meaningful
distinction between the low scale models and the SM in 5 - 10 years of
data.  Using the WB bound on the optically thin source flux, we find
that the upward events are too sparse to discriminate among models by
use of the R ratio.  Flux values in between the two presented here
offer various levels of discrimination, with useful information
obtainable for the larger fluxes.
 
Putting together the rise in downward event rate above 1 PeV, the
sensitivity of R to the new physics cross sections, and the slope of the
upward flux as a function of nadir angle, we see that a signal of
low scale gravity models would stand out clearly.

\section{Conclusions:} $KM^3$ detectors have been planned primarily as
``neutrino telescopes''.  However our study indicates that $KM^3$ will
also provide vigorous new exploration of fundamental questions in
ultra-high energy physics.  Extra dimension, low scale gravity models
show an observable impact on the signature of neutrino events in the
$KM^3$ detectors such as the currently running RICE detector and the
ICECUBE detector, which is in the R \& D stage.  Of the ``strong
gravity'' effects we looked at, the black hole formation cross section
and the extrapolation of the small $q^{2}$, part of $\sigma^{\nu N}$
from the $s < M^2$ to $ s \gg M^2$ within the ADD model yield the
largest detectable effects.

While some observables are somewhat sensitive to the number of
dimensions, $n$, most are {\it quite sensitive} to the value of the
scale of gravity $M$.  If $M$ is 1-2 TeV, the enhanced $\sigma^{\nu
N}$ creates a clearly recognizable signature, with the ``neutral-to-
charged'' event ratio still showing new physics effects at M = 5 TeV.
The ratio $R$ of upcoming to down-going events is a powerful
diagnostic, capable of discriminating between models if fluxes are
large enough to produce a significant number of upcoming events. In
the entire analysis, we emphasized that the regeneration of neutrino
upcoming flux due to neutrinos scattering down from higher to lower
energies is a crucial feature of the gravity-induced neutral current
interactions.  This is a general and important feature of our work
presented here.

Considering only down-coming events, we propose that the fraction of
neutral current type events, those with hadronic showers and no
leading charged lepton, provides a probe of new interactions.  In
particular, event signatures arising from a {\it dominant} component
of eikonalized graviton exchange or of black hole production as occur
in low scale gravity models, would have {\it no standard physics
explanation}, and would point the way toward physics beyond the
Standard Model if observed.  There is every reason to believe that the
neutral-to-charged current ratio can be extracted from upcoming
facilities well enough to make a practical signal.  In particular,
determining the ratio of events with a muon, to those making an
isolated shower, is certainly feasible with a combination of
AMANDA/ICECUBE and RICE technology.  The striking behavior of the
neutral current-to-charged current ratio is shown in Fig. \ref{nc/cc}.

Finally, the shape and {\it slope} of the angular distribution, Fig.
\ref {flux_ang}, is found to have good discriminating power.  The
shape does not depend at all on the overall normalization of the flux.
Moreover the slopes differ substantially right in the regime of maximum
detectable flux, near $\pi/2$ nadir angle, and is ideal for comparing low scale
gravity models to the Standard Model.  Whether or not extra-dimension
models as currently envisaged survive, the angular distribution can severely
test the neutrino physics of the Standard Model, possibly strongly bounding or 
even discovering new physics, as soon as data becomes available.

\noindent
{\bf Acknowledgements:} P. Jain thanks the University of Kansas 
College of Arts and Sciences and Department of Physics and Astronomy 
for hospitality and support in the course of this work.  This research
was supported in part by the U.S. Department of Energy under grant
number DE-FG03-98ER41079.  We used computational facilities of the Kansas
Center for Advanced Scientific Computing for part of this work.

\end{document}